\documentclass[preprint,aps,tightenlines,showpacs,nofootinbib]{revtex4}
\usepackage{latexsym}
\usepackage[dvips]{graphicx}
\newcommand{\beq}{\begin{eqnarray}}
\newcommand{\eeq}{\end{eqnarray}}
\newcommand{\eq}{eqnarray}

\newcommand{\ci}{\cite}

\newcommand{\Om}{{\Omega}}

\newcommand{\no}{{\nonumber}}
\newcommand{\f}{\frac}

\newcommand{\Sch}{Schwarzschild }

\begin{document}

\preprint{hep-th/0405045}

\title{The Final State of Black Strings and $p$-Branes, and
    the Gregory-Laflamme Instability
}

\author{Mu-In Park\footnote{E-mail address: muinpark@yahoo.com}}

\affiliation{ Department of Physics, POSTECH, Pohang 790-784,
Korea}

\begin{abstract}
It is
%CGQG<
%widely believed
shown
%CGQG>
that the usual entropy argument for the Gregory-Laflamme (GL)
instability for $some$ appropriate black strings and $p$-branes
%CGQG<
gives  surprising agreement up to a few percent.
%indicates only some plausibility but should not be taken
%seriously. But I show that this widespread belief is not correct:
%Correct application of the entropy argument gives quite good
%agreement with the linearized stability analysis up to $0.5 \sim
%2.4~ \%$ discrepancy. This agreement indicates that there might
%exist a deep connection between the global entropy argument and
%the classical instability, similarly to Gubser-Mitra's case. On
%the other hand,
This may provide
%also
a strong support to the GL's horizon fragmentation, which would
produce the array of higher-dimensional Schwarzschild-type's black
holes finally. On the other hand, another estimator for the size
of the black hole end-state relative to the compact dimension
indicates a second order (i.e., smooth) phase transition for some
$other$ appropriate compactifications and total dimension of
spacetime wherein the entropy argument is not appropriate.
%:
%For the black strings $Sch_n \times L$, this occurs for $n$ as
%large as $n>12$ and for the (isotropic) black $p$-branes $Sch_n
%\times L^p$, this occurs for $n$ as low as $n<6$ with a fixed
%total dimension of spacetime $d=10$.
In this case,
Horowitz-Maeda-type's non-uniform black strings or $p$-branes can
be the final state of the GL instability.
%is disproved up to this
%discrepancy.
%And if the entropy argument is correct, the discrepancy would be,
%probably, partly due to the crude approximation for the final
%states and partly due to some possible uncertainty in the
%numerical analysis.

\end{abstract}


\maketitle

\newpage

\section{Introduction}

The four-dimensional \Sch black hole in Einstein gravity is
well-known to be stable classically under linearized perturbations
\ci{Chan:83}. Recently, it has been shown that this extends to
hold for higher dimensional cases \ci{Ishi:03}. However, Gregory
and Laflamme discovered that the black strings and $p$-branes of
10-d low energy string theory, which have {\it hypercylindrical}
horizons $Sch_n \times V_p$ instead of compact {\it
hyperspherical} ones $Sch_{n+p}$, are found to be unstable as the
compactification scale, say $L$, of extended directions becomes
larger than the order of the horizon radius $r_+$--the so-called
Gregory-Laflamme (GL) instability \ci{Greg:93}. In GL's original
work, they explained the instability by arguing that a black
string $Sch_4 \times L$ has a lower entropy than a 5-d \Sch black
hole $Sch_5$ with the same total mass when $L > r_+$, in the
context of microcanonical ensemble \footnote{Another explanation,
based on $D-\bar{D}$ pair annihilations, is also known though it
gives only the order of magnitudes \ci{Dani:01}.
%; I thank U. H. Danielsson for the information.
}; and they also argued that this lend support to the horizon
fragmentation, which would produce array of black holes
eventually. However, it is widely believed that this entropy
argument for the classical stability should not be taken seriously
since it
%CGQG<
$estimates$
%predicts
%CGQG>
a wrong onset point of the instability--this
means the black string can be classically stable even if its
entropy is smaller than that of 5-d \Sch black hole for some
regime of $L$--though it provides some plausibility argument
\ci{Gubs:00,Real:01,Hira:01,Hube:02}. Moreover, the GL's
fragmentation scenario was disproved under very weak assumptions,
including the $classical$ black hole area theorem, by Horowitz and
Maeda (HM) and a non-uniform black string as the final state of
the GL instability \ci{Horo:01} is considered accordingly.

In this paper, I will show that this widespread belief is $not$
%CGQG<
$quite$
%CGQG>
correct: If one properly apply the entropy argument to the black
string solution $Sch_9 \times L$ of 10-d low energy string theory,
one can
%CGQG<
estimate
%get
%CGQG>
the
%CGQG<
%correct
%CGQG>
onset point of
%CGQG<
the GL
%a classical
%CGQG>
instability
%CGQG<
%which agrees with GL's numerical analysis
%CGQG>
up to 2.4 \% discrepancy.
%HM's non-uniform black string state as the final
%state is disproved up to this discrepancy.
For $p$-brane solutions, the thing depends on the geometry of the
compactification of $p$-branes. I consider two typical methods of
compactifications: Thin-torus compactification and $p$-dimensional
isotropic-torus compactification. For the former case, the
discrepancy grows as $p$ grows ($n$ decreases) up to 35 \%
discrepancy for $p=6$ ($n=4$). But, for the latter case, the
discrepancy is quite reduced up to $0.5 \sim 2.4~ \%$.
%CGQG<
This may provide a strong support to the GL's horizon
fragmentation, which would produce the array of higher-dimensional
Schwarzschild-type's black holes finally. On the other hand,
another estimator for the size of the black hole end-state
relative to the compact dimension indicates a second order (i.e.,
smooth) phase transition for some $other$ appropriate
compactifications and total dimension of spacetime wherein the
entropy argument is not appropriate: For the black strings $Sch_n
\times L$, this occurs for $n$ as large as $n>12$ and for the
(isotropic) black $p$-branes $Sch_n \times L^p$, this occurs for
$n$ as low as $n<6$ with a fixed total dimension of spacetime
$d=10$. In this case, HM-type's non-uniform black strings or
$p$-branes can be the final state of the GL instability instead.
%If the
%entropy argument is correct, the discrepancy would be, probably,
%partly due to the crude approximation, that I take, for the final
%states and the idealized compactification of $p$-branes.
%CGQG>

\section{The black string instability}
\label{2}

The black string and $p$-branes I am specifically interested in
are those introduced by Horowitz and Strominger \ci{Horo:91} in
10-d low energy string theory with a metric given by
\begin{\eq}
ds^2=-N^2 dt^2 +N^{-2} dr^2 + r^2 d \Om_{n-2}^2 +dx^i dx_i,
\end{\eq}
where
\begin{\eq}
 N^2=1-\f{16\pi G M_{(n)}}{(n-2) \Om_{n-2}r^{n-3}},
\end{\eq}
$n=4, \dots, 10$ and index $i$ runs from 1 to $p=10-n$. $M_{(n)}$
is the mass of the $n$-dimensional black holes, and $ \Om_{n-2}$
is the area of the unit sphere $S^{n-2}$ \ci{Myer:86}. This is
always a solution of the Einstein equation in $d=n+p=10$
dimensions for compact as well as non-compact string or brane
directions if the string or brane directions are completely
factorized. But, this particular solution does not exist for
$0$-brane (i.e., 10-d black hole) and we must consider
deformations of ordinary \Sch solution due to non-compact
dimensions in general \ci{Hube:02,Horo:96,Myer:87}. But, let me
approximate the 10-d black hole by the ordinary $Sch_{10}$ metric,
with 10-d radial coordinate $R$.

In order to compute the transition point between the black strings
or branes and the 10-d black hole of the same mass due to the
entropy difference, in the context of microcanonical ensemble, we
need to know details of compactified dimensions. In this section
let me first consider the simplest one, black string and consider
simply a $S^1$-compactification. To this end, let me note that the
masses and entropies of the black string and the 10-d black hole,
with the horizon radii $r_+$ and $R_+$, are, respectively
\begin{\eq}
&& M_{b.s.}=\f{7 \pi^3 r_+ L}{48 G}, ~~ S_{b.s.}=\f{\pi^4 r_+^7
L}{12 G}, \no \\
&& M_{b.h.(10)}=\f{16 \pi^3 R_+^7 }{105 G}, ~~ S_{b.h.(10)}=\f{8
\pi^4 R_+^8 }{105 G} \no.
\end{\eq}

Now for the same mass of the black string and the 10-d black hole,
the condition of an unstable black string due to the smaller
entropy than 10-d black hole is
\begin{\eq}
\label{L:thin} L \geq \left(\f{8}{7} \right)^8
\left(\f{\Om_8}{\Om_7} \right) r_+ \approx 2.661 ~r_+.
\end{\eq}
Note also that
\begin{\eq}
L \geq \left(\f{8}{7} \right)^7 \left(\f{\Om_8}{\Om_7} \right) R_+
\approx 2.328 ~R_+,
\end{\eq}
such as the 10-d black hole can easily fit in the compact
dimension $S^1$. In terms of the wave number $k$ for the unstable
perturbation \ci{Gubs:02}, (\ref{L:thin}) can be re-expressed  as
\begin{\eq}
k \leq k_{S},~k_{S} \equiv \f{2 \pi}{L_{S}} \approx
2.361~r_+^{-1},
\end{\eq}
where $L_{S}$ is the entropy estimator--``equal entropy for equal
mass'' estimator--of the minimum length of compact dimension for
the GL instability. This agrees with the GL's numerical analysis
for the classical instability under linearized perturbations $k
\leq  k_{GL},~k_{GL} \approx 2.306~ r_+^{-1}$ up to 2.4 \%
discrepancy \footnote{$2 \mu$ in GL's analysis is what I have
called $k$ \ci{Gubs:02}. And I use values of $\mu$ recently
obtained by Hirayama et al. \ci{Hira:03}, which is more accurate
than the original GL's analysis; I thank G. Kang for informing
about this updated data. Similar data has been also obtained by E.
Sorkin \ci{Sork:04,Sork:priv} in a different context of Gubser
\ci{Gubs:02} and Wiseman \ci{Wise:02}; I thank E. Sorkin for
kindly sending his data.}. This good agreement is rather
surprising since thermodynamic instability based on $global$
entropy arguments, which have quantum origins, does not generally
imply a classical instability.
%CGQG<
%; but, this might indicate a deep
%connection between the global entropy argument and the classical
%instability, similarly to Gubser-Mitra's case
%\ci{Gubs:00,Real:01,Hira:01,Hube:02,Greg:01,Kang:04}. So, I
%$assume$ the entropy argument be correct, as a {\it working
%hypothesis} hereafter. Now, if the entropy argument, which depends
%crucially on the choice of the entropy-maximum state, is correct,
%the small discrepancy would imply then that my crude approximation
%of ordinary $Sch_{10}$ metric as
%%the 10-d black hole solution
%the final state of the black string $Sch_9 \times L$, even with a
%compact dimension $S^1$, is quite good;
%%On the other hand, this
%%also implies that is the $Sch_{10}$ solution and HM's non-uniform
%%black string state as the final state may be disproved up to that
%%discrepancy
%deformation, due to a compact dimension, of the final state from
%$Sch_{10}$ would be only a few percents and moreover the actual
%deformation would be somewhat smaller if one notes some possible
%uncertainty in the GL's numerical analysis \ci{Greg:93,
%Greg:94,Hira:03}.
%CGQG>

\section{The black $p$-brane instability I: Thin-torus compactification}
\label{3} The generalization of the string instability of the
previous section to arbitrary $p$-branes ($2 \leq p \leq 6$) in
10-d low energy string theory requires the knowledge on the
compactfication.  In this section, I first consider a {\it
thin-torus} compactification with horizons $Sch_n \times L \times
V_{p-1}$ which has one compact dimension $S^1$ with length $L$ and
a very tiny volume $V_{p-1} \ll L^{p-1}$ for other compact
dimensions. Since the effect  of small compact dimensions would be
tiny, I would approximate this system by the black strings $Sch_n
\times L $ in $(n+1)$-dimensions effectively such as the
transition problems between $p$-branes and 10-d black holes are
reduced to those of black strings $Sch_n \times L $ and $(n+1)$-d
black holes. To this end, similarly to the previous section, let
me approximate the $(n+1)$-d black holes by the ordinary
$Sch_{n+1}$ metric, with $(n+1)$-d radial coordinate $R$. Then,
the masses and entropies of the black string $Sch_n \times L $ and
the $(n+1)$-d black hole are, respectively,
\begin{\eq}
&& M_{b.s.(n)}=\f{(n-2) \Omega_{n-2} r_+^{n-3} L}{16 \pi G}, ~~
S_{b.s.(n)}=\f{\Omega_{(n-2)} r_+^{n-2}
L}{4 G},\no \\
&& M_{b.h.(n+1)}=\f{(n-1) \Omega_{(n-1)} R_+^{n-2} }{16 \pi G}, ~~
S_{b.h.(n+1)}=\f{\Omega_{n-1} R_+^{n-1} }{4 G} \no.
\end{\eq}

Now, for the same mass of the black string and the $(n+1)$-d black
hole, the condition of an unstable black string due to the smaller
entropy than $(n+1)$-d black hole is
\begin{\eq}
\label{k:Thin}
 L \geq \left(\f{n-1}{n-2} \right)^{n-1}
\f{\Om_{n-1}}{\Om_{n-2}} r_+ \equiv 2 \pi ~k_{S}^{-1},
\end{\eq}
where $k_{S}$ is the entropy estimator of the maximum wave number
for an unstable perturbation. Note also that
\begin{\eq}
\label{f:Thin}
 L \geq \left(\f{n-1}{n-2} \right)^{n-2}
\f{\Om_{n-1}}{\Om_{n-2}} R_+ \equiv 2 f ~R_+,
\end{\eq}
where $f=(\f{n-1}{n-2})^{n-2} \Omega_{n-1}/2 \Omega_{n-2}$ denotes
how $(n+1)$-d black hole can fit in the compact direction $S^1$,
such as $f \geq 1$ is required for a safe fitting. The values
computed for $k_{S}$ and $f$ are listed and compared with the GL's
data $k_{GL}$ in Table I, Fig. 1 and Table II, Fig. 2,
respectively. Table I and Fig. 1 show that the discrepancy grows
as $p$ grows ($n$ decreases) up to 35 \% discrepancy for
$p=6~(n=4)$. But, in the light of entropy argument
%CGQG<
%, a working hypothesis,
these discrepancy would only imply that the crude
approximation, that I have taken, for the $Sch_{(n+1)}$ metric as
the $(n+1)$-d black hole solution even with one compactified
dimension $S^1$, and/or the thin torus limit of the
compactification, which treats one specific direction differently
from others, becomes bad as the dimension of the compactification
$p$ increases. So, this indicates that better approximation which
treats equally all the compact directions is needed. This will be
done in the next section. But, before this, let me note the
followings.
\begin{table}
\begin{center}
\begin{tabular}{|c|c|c|c|} \hline
n& GL's data  & Isotropic Torus & Thin Torus\\
\hline \hline 4 & 0.880 & 0.857 (-3 \%) & 1.185 (+35 \%) \\ \hline
5 & 1.27 & 1.206 (-5 \%) & 1.491 (+17 \%) \\ \hline 6 & 1.58 &
1.524 (-4 \%) & 1.748
(+11 \%) \\ \hline 7 & 1.85 & 1.820 (-2 \%) & 1.973 (+7 \%) \\
\hline 8 & 2.088 & 2.098 (+0.5 \%) & 2.176 (+4.2 \%) \\ \hline 9 &
2.306 & 2.361 (+2.4 \%) & 2.361 (+2.4 \%) \\ \hline
\end{tabular}
\end{center}
\caption{Table of the entropy estimator $k_{S}$ for isotropic
torus and thin torus compactifications in comparison with the GL's
data $k_{GL}$. The values in the brackets denote their
discrepancies to GL's data ($r_+ \equiv 1$); the $+$ or $-$ sign
represents whether it is bigger ($+$) or smaller ($-$) than GL's.}
\end{table}
\begin{figure}
\includegraphics[width=10cm,keepaspectratio]{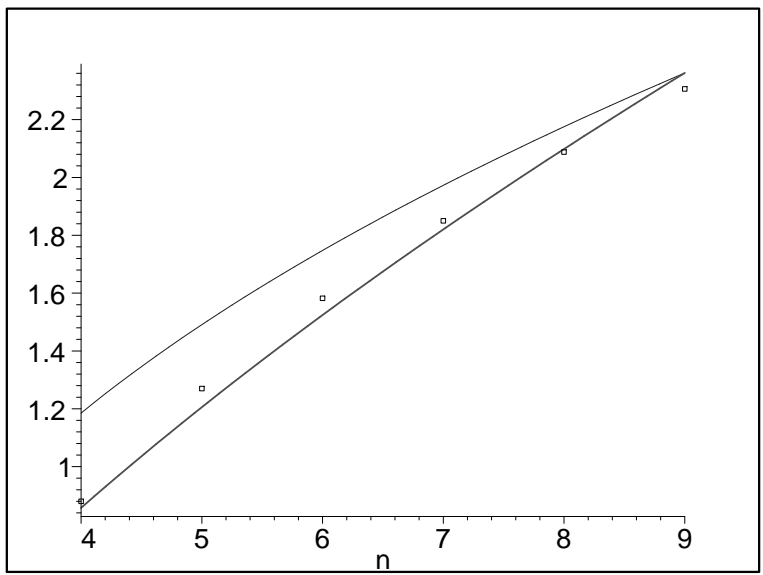}
\caption{Plot of $k_{S}$ as a function of $n$ for black string
$(n=9)$ and $p$-branes ($n=10-p$). The boxed points represent GL's
numerical data $k_{GL}$, the thin and the thick lines represent
the values calculated for thin torus and isotropic torus
compactifications, respectively.} \label{fig:kmax}
\end{figure}
\begin{table}
\begin{center}
\begin{tabular}{|c|c||c|c|} \hline
n&  Isotropic Torus &n& Thin Torus\\ \hline \hline 4 & 0.916 & 8 &
1.238
 \\ \hline 5 & 0.977 * & 9 & 1.164 \\
\hline 6 & 1.031 * & 10 & 1.102
 \\ \hline 7 & 1.079 & 11 & 1.049 \\
\hline 8 & 1.123 & 12 & 1.003 * \\ \hline 9 & 1.164 & 13 & 0.962 *\\
\hline
\end{tabular}
\end{center}
\caption{Comparison of $f$ for isotropic torus and thin torus. The
marked (*) ones are the two nearest dimensions to the critical
dimension $n_c$.}
\end{table}
\begin{figure}
\includegraphics[width=10cm,keepaspectratio]{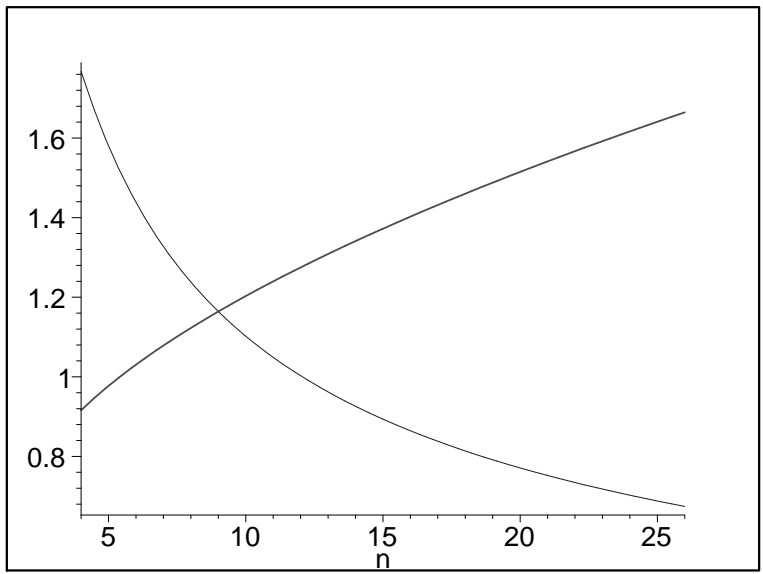}
\caption{Plot of $f$ as a function of $n$ for black string $(n=9)$
and $p$-branes $(n=10-p)$. The thin and the thick lines represent
the values calculated for thin torus and isotropic torus
compactifications, respectively. The crossing occurs at about
$n=9$. } \label{fig:f}
\end{figure}

First, the widespread belief \ci{Gubs:00,Real:01,Hira:01,Hube:02}
that the entropy argument for the classical instability should not
be taken seriously was originated from  the big discrepancy of 35
\% with GL's numerical analysis for the $n=4$ case, which is found
to be the worst case in the thin torus, i.e., string,
approximation of the $p$-branes of the 10-d low energy string
theory. But my analysis shows that this is not quite correct,
since $n=4$ case is not truly a black string but a 6-brane exactly
\footnote{This is sharply contrast to the equations for the linear
perturbations \ci{Greg:93}, which depend only on the sum of the
Kaluza-Klein mass-squared, i.e., $\mu^2=\sum_{i=1}^p \mu^2_i$ and
is blind to the dimensionality $p$ of the brane's world volumes as
long as $\mu^2 \neq 0$.}, which would deform the black string
picture quite much; one should have considered the $n=9$ case to
discuss the black string and compare with GL's data.

Second, note that the result (\ref{k:Thin}) and (\ref{f:Thin}) can
be applied for any dimension $n$ to analyze the transition from a
black string $Sch_n \times L$ to $(n+1)$-d black hole $Sch_{n+1}$,
though I have introduced this set-up to approximate the $p$-brane
solutions in 10-d string theory. Then, it is interesting to
observe that there is a {\it critical dimension} $n_c=12$ above
which $f <1$ such as $(n+1)$-d black hole can not fit in the
compact direction $S^1$; in this case, approximating the ordinary
$Sch_{(n+1)}$ as the final state solution needs some
%CGQG<
important
%CGQG>
correction due to the compact dimension \ci{Myer:87}
%CGQG<
such as the black string can evolve into a different final state,
presumably a non-uniform black string, between the (uniform) black
string and the black hole. This indicates that the order of the
phase transition between the uniform and the non-uniform black
strings changes from the first (i.e., sudden transition)  to
second order (i.e., smooth transition) at the critical dimension
$n_c$.
%CGQG>
Recently another estimator for the critical dimension has been
considered by Sorkin \ci{Sork:04} but one finds a very good
agreement between these two estimators.

\section{The black $p$-brane instability II: isotropic $p$-dimensional torus compactification}

As a correction to the thin-torus compactification of the previous
section, I will consider an isotropic $p$-dimensional torus
compactification $Sch_n \times V_p$ where all compactified
directions are treated equally. To this end, let me approximate
10-d black hole by the ordinary $Sch_{10}$ metric, with 10-d black
radial coordinate $R$, similarly to Sec. \ref{2}. Then, the masses
and entropies of a black $p$-brane and 10-d black hole are,
respectively,
\begin{\eq}
&& M_{b.b.(n)}=\f{(n-2) \Omega_{n-2} r_+^{n-3} V_p}{16 \pi G}, ~~
S_{b.b.(n)}=\f{\Omega_{(n-2)} r_+^{n-2}
V_p}{4 G},\no \\
&& M_{b.h.(10)}=\f{16 \pi^3 R_+^7 }{105 G}, ~~ S_{b.h.(10)}=\f{8
\pi^4 R_+^8 }{105 G} \no.
\end{\eq}

Now, for the same mass of the black $p$-branes and the 10-d black
hole, the condition of an unstable black $p$-branes due to the
smaller entropy than 10-d black hole is
\begin{\eq}
\label{Vp1:iso}
 V_p \geq \left(\f{8}{n-2} \right)^8
\f{\Om_8}{\Om_{n-2}} r_+^{10-n}
\end{\eq}
while
\begin{\eq}
\label{Vp2:iso}
 V_p \geq \left(\f{8}{n-2} \right)^{n-2}
\f{\Om_8}{\Om_{n-2}}  R_+^{10-n}.
\end{\eq}
Moreover, since I am considering a $p$-dimensional torus with
equal length $L =(V_p)^{1/p}$, (\ref{Vp1:iso}) and (\ref{Vp2:iso})
can be re-expressed, in terms of $L$ and the associated entropy
estimator of the maximum wave number $k_{S}$ for the unstable
perturbation, as
\begin{\eq}
\label{k:iso}
 L \geq \left(\f{8}{n-2} \right)^{\f{8}{10-n}}
\left(\f{\Om_{8}}{\Om_{n-2}}\right)^{\f{1}{10-n}} r_+ \equiv 2 \pi
~k_{S}^{-1}
\end{\eq}
and
\begin{\eq}
\label{f:iso}
 L \geq \left(\f{8}{n-2} \right)^{\f{8}{10-n}}
\left(\f{\Om_{8}}{\Om_{n-2}}\right)^{\f{1}{10-n}} R_+ \equiv 2 f
~R_+.
\end{\eq}
The values computed for $k_{S}$ and $f$ are listed and plotted
also in Table I, Fig. 1 and Table II, Fig. 2, respectively, in
comparison with GL's data $k_{GL}$ and the results for the
thin-torus compactfication. Table I and Fig. 1 show that the
discrepancy of the thin torus approximation have been quite
reduced in the isotropic torus and the worst one is about $4 \sim
5$ \% for $p=4,5~(n=6,5)$; all the other cases have been less than
about $2 \sim 3~ \%$ and the best one is 0.5 \% for $p=2~(n=8)$;
moreover, compared to the growing discrepancy for the thin-torus
as $p$ grows ($n$ decreases), that of the isotropic torus is
almost stable such as this improved approximation is fairly good.

On the other hand, according to the result for $f$ in Table II and
Fig. 2, there is a critical dimension $n_c=6$ below which $f < 1$
such as the 10-d black hole can not fit in the compact dimension
$S^1$.
%CGQG<
This implies that approximating the ordinary $Sch_{10}$ as the
final state of the black $p$-branes
%the full 10-d black hole solution
needs some important corrections due to the compactified dimension
such as the uniform black $p$-brane can evolve into a different
final state, presumably a non-uniform black $p$-brane. So, the
relatively big discrepancies for $n=5, 6$ would not be so
surprising in the light of entropy argument; but it is a
remarkable fact that $n=4$ case has a relatively good agreement
with GL's data with 3 \% discrepancy even though it does not have
to be.
%, which being a mystery.
Hence, by taking into account this additional fact to the result
of $k_{S}$ in Table I and Fig. 1, the true discrepancy in this
approximation would be quite smaller and the reliable results
would have discrepancy only about 0.5 \% $\sim$ 2.4 \% by
excluding $n=4,5,6$ cases.

Furthermore, this also indicates a smooth decay of an unstable
(uniform) black $p$-brane $Sch_n \times L^p$ to the non-uniform
state for $n$ as low as $4$ or $5$. This is in contrast to the
decay of black string, where $n$ as large as $n>12$ is required
for a smooth decay. More recently another estimator for the
critical dimension, following Sorkin \ci{Sork:04}, has been also
considered by Kol and Sorkin \ci{Kol:04} and they found a very
good agreement with mine again; moreover they found interestingly
that $d=10$ is the smallest total dimension of the spacetime to
allow a smooth decay of an unstable black brane to the non-uniform
state.

%On the other hand this also implies
%that the final state of the black $p$-branes is the $Sch_{10}$
%solution with less than 2.4 \% discrepancy and the HM-type's
%non-uniform black $p$-brane states as the final state may be also
%disproved up to that discrepancy.

\section{Discussion}

I have shown that
%CGQG<
the usual entropy argument for the GL instability for $some$
appropriate black strings and $p$-branes
%CGQG>
gives  surprising agreement up to a few percent. This may provide
a strong support to the GL's horizon fragmentation, which would
produce the array of--single in my analysis--higher dimensional
Schwarzschild-type's black holes finally; this result is
remarkable in that the end point of the unstable evolution, which
is by its nature ``non-linear'', crucially affects the onset of
the instability calculation, which is by its nature ``linear''.

%CGQG<
On the other hand, another estimator for the size of the black
hole end-state relative to the compact dimension indicates a
second order (i.e., smooth) phase transition for some $other$
appropriate compactifications and total dimension of spacetime
wherein the entropy argument is not appropriate. For the black
strings $Sch_n \times L$, this occurs for $n$ as large as $n>12$
and for the (isotropic) black $p$-branes $Sch_n \times L^p$, this
occurs for $n$ as low as $n<6$ with a fixed total dimension of
spacetime $d=10$. In this case, HM-type's non-uniform black
strings or $p$-branes can be a natural final state of the GL
instability. This result agrees quite well with the analysis of
Kol and Sorkin wherein different estimator has be considered
\cite{Sork:04,Kol:04}.

\begin{figure}
\includegraphics[width=10cm,keepaspectratio]{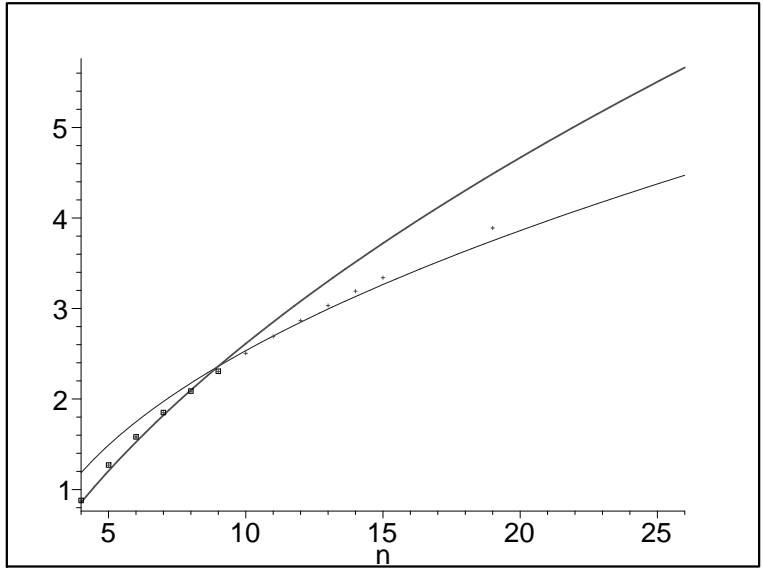}
\caption{Plot of $k_{S}$ as a function of $n$ for Sorkin's black
string analysis in arbitrary dimensions, in comparison with Fig.1;
cross points correspond to Sorkin's data
%--$
%k_min_Sorkin_6_brane_1 :=
%0.876, % (-2 \%),
%k_min_Sorkin_5_brane_1:=
%1.269, % (-5.0 \%),
% k_min_Sorkin_4_brane_1 :=
%1.581, % (-3.6 \%),
%k_min_Sorkin_3_brane_1 :=
%1.849, % (-1.5 \%),
% k_min_Sorkin_2_brane_1 :=
%2.087, % (+0.5 \%),
% k_min_Sorkin_1_brane_1 :=
%2.30, % (+3 \%),
% k_min_Sorkin_10_string_1 :=
%2.50, % (+1 \%),
% k_min_Sorkin_11_string_1 :=
%2.691, % (+0.2 \%),
%k_min_Sorkin_12_string_1 :=
%2.867, % (-0.6 \%),
% k_min_Sorkin_13_string_1 :=
%3.033, % (-1.3 \%),
% k_min_Sorkin_14_string_1 :=
%3.191, % (-1.8 \%),
% k_min_Sorkin_15_string_1 :=
%3.342, % (-2.3 \%),
% k_min_Sorkin_19_string_1 :=
%3.889, % (-3.6 \%),
% k_min_Sorkin_29_string_1 :=
%5.060, % ( -6.2 \%),
% k_min_Sorkin_49_string_1 :=
%6.820$% (-7.8 \%)
%~for $n=4, \dots, 15, 19, 29, 49$ --
\ci{Sork:04,Sork:priv} and
these overlap with GL's data for $n \leq 9$ .} \label{fig:Sorkin}
\end{figure}

{\it Note added}: After the first appearance of this paper, I was
informed by E. Sorkin that my analysis of the instability for
thin-torus compactification is very similar to that of Ref.
\ci{Sork:04} which uses a single dimensionless parameter
$\tilde{\mu} \equiv G M/L^{n-2}$ instead of $r_+, R_+, L$.
Afterward, I have checked that his result (9) on the critical
values of $\tilde{\mu}$ for the onset of an instability agrees
exactly with my result (6). And his analysis
\ci{Sork:04,Sork:priv} on the black string perturbation
%, following Gubser \ci{Gubs:02} and Wiseman \ci{Wise:02},
shows a quite good agreement with the thin-torus set-up up to $0.2
\sim 1 \% $ discrepancy for $10 \leq n \leq 12$, in contrast to $n
\leq 9$, where the isotropic-torus set-up is more favorable ( Fig.
\ref{fig:Sorkin}); this indicates {\it another critical dimension}
at $n=9$ which agrees with Kol's ``merger point'', where the
string and black hole branches merge  \ci{Kol:02}. The increasing
discrepancy above $n=12$ is not so surprising since this is the
regime where the naive thin-torus set-up does not have to be
correct, due to $f<1$, such as the black string can evolve into a
different final state, presumably a non-uniform black string,
%such as some correction due to compact dimension is necessary
as in $ n < 6$ cases of isotropic-torus set-up in Sec. IV.
%This provides a strong evidence on the entropy argument of the black
%string instability.
%He also considered the
%effect of the leading order correction to the higher dimensional
%\Sch metric due to a $S^1$ compactification, following Harmark's
%solution \ci{Harm:03}, and have found (corrected Fig. 1 of Ref.
%\ci{Sork:04}) an enhancement of the discrepancy in the thin-torus
%compactification. However, it does not necessarily mean a
%contradiction to my expectation in Sec. \ref{3} since the
%Harmark's solution \ci{Harm:03} does not seems to be fully
%consistent with his numerical data (Fig. 2 of Ref. \ci{Sork:04}),
%as well as others \ci{Gubs:02,Wise:02}, which shows that entropy
%correction should be negative, below $n=12$.

\section*{Acknowledgments}

I would like to thank  Ulf Danielsson, James Gregory, Steven
Gubser, Troels Harmark, Gary Horowitz, Akihiro Ishibashi, Gungwon
Kang, Hideo Kodama, Luis Lehner, and Evgeny Sorkin for useful
correspondences. I was supported by the Korea Research Foundation
Grant (KRF-2002-070-C00022).

%%%%%%%%%% References %%%%%%%%%%%%%%%%%%%%%%%%%
\newcommand{\J}[4]{#1 {\bf #2} #3 (#4)}
\newcommand{\andJ}[3]{{\bf #1} (#2) #3}
\newcommand{\AP}{Ann. Phys. (N.Y.)}
\newcommand{\MPL}{Mod. Phys. Lett.}
\newcommand{\NP}{Nucl. Phys.}
\newcommand{\PL}{Phys. Lett.}
\newcommand{\PR}{Phys. Rev. D}
\newcommand{\PRL}{Phys. Rev. Lett.}
\newcommand{\PTP}{Prog. Theor. Phys.}
\newcommand{\hep}[1]{ hep-th/{#1}}
\newcommand{\hepp}[1]{ hep-ph/{#1}}
\newcommand{\hepg}[1]{ gr-qc/{#1}}
%%%%%%%%%%%%%%%%%%%%%%%%%%%%%%%%%%%%%%%%%%%%%%%

\end{document}